\documentclass{ptephy_v1}

\preprintnumber{XXXX-XXXX} 

\usepackage{multirow}

\begin{document}

\title{Direction-sensitive dark matter search with a low-background gaseous detector NEWAGE-0.3b''}


\author{Tomonori Ikeda}
\affil{Department of Physics, Graduate School of Science, Kobe University, Rokkodai-cho, Nada-ku, Kobe-shi, Hyogo, 657-8501, Japan \email{ikeda.tomonori.2s@kyoto-u.ac.jp}}
\author[1,2]{Kiseki Nakamura}
\affil{Kamioka Observatory, Institute for Cosmic Ray Research, the University of Tokyo, Higashi-Mozumi, Kamioka-cho, Hida-shi, Gifu, 506-1205, Japan}
\author[1]{Takuya Shimada}
\author[1]{Ryota Yakabe}
\author[1]{Takashi Hashimoto}
\author[1]{Hirohisa Ishiura}
\author[1]{Takuma Nakamura}

\author[2,4]{Hiroshi Ito}
\author[3]{Koichi Ichimura}
\affil{Research Center for Neutrino Science, Tohoku University, Sendai 980-8578, Japan}
\author[2,4]{Ko Abe}
\affil{Kavli Institute for the Physics and Mathematics of the Universe (WPI), the University of Tokyo, 5-1-5 Kashiwanoha, Kashiwa-shi, Chiba, 277-8582, Japan}
\author[5]{Kazuyoshi Kobayashi}
\affil{Waseda Research Institute for Science and Engineering, Waseda University, 3-4-1 Okubo, Shinjuku, Tokyo 169-8555, Japan}
\author[6]{Toru Tanimori}
\affil{Division of Physics and Astronomy, Graduate School of Science, Kyoto University, Kitashirakawaoiwake-cho, Sakyo-ku, Kyoto-shi, Kyoto, 606-8502, Japan}
\author[6]{Hidetoshi Kubo}
\author[6]{Atsushi Takada}
\author[2,4]{Hiroyuki Sekiya}
\author[2,4]{Atsushi Takeda}
\author[1]{Kentaro Miuchi}


\begin{abstract}%
NEWAGE is a direction-sensitive dark matter search using a low-pressure gaseous time projection chamber.
A low alpha-ray emission rate micro pixel chamber had been developed in order to reduce background for dark matter search.
We conducted the dark matter search at the Kamioka Observatory in 2018.
The total live time was 107.6~days corresponding to an exposure of 1.1~kg$\cdot$days.
Two events remained in the energy region of 50-60~keV which was consistent with 2.5 events of the expected background. A directional analysis was carried out and no significant forward-backward asymmetry derived from the WIMP-nucleus elastic scatterings was found. Thus a 90\% confidence level upper limit on Spin-Dependent WIMP-proton cross section of 50~pb for a WIMP mass of 100~GeV/$c^2$ was derived.
This limit is the most stringent yet obtained from direction-sensitive dark matter search experiments.
\end{abstract}

\subjectindex{Dark matter, MPGD, $\mu$TPC}

\maketitle

\section{Introduction}
Dark matter is one of the biggest puzzles of the modern cosmology and the particle physics. 
A number of experimental efforts aiming to find the Weakly Interacting Massive Particle (WIMP) dark matter through direct searches which observe the scatterings of the WIMP and nuclei have been carried out~\cite{XENON,LUX,CDMS,DarkSide,DEAP3600,EDELWEISS,PANDAX,XMASS}. 
However, the dark matter has not been discovered yet.
In the direct search, the annual modulation and the directional signature would be two possible signals 
among the characteristic signals of the dark matter.
The annual modulation is caused by the orbital motion of the Earth around the Sun. The modulation amplitude is expected to be a few~\%~\cite{BAUM2019262}.
On the other hand, the directional signature is due to the circular motion of the solar system around the galaxy center. The forward-backward ratio in nuclear recoil angular distribution derived from the WIMP-nucleus elastic scatterings could be an order of magnitude~\cite{Spergel}. 
In addition, the directional method could discover the WIMP dark matter beyond the neutrino floor, which represents the ultimate background by the coherent neutrino-nucleus scatterings~\cite{PhysRevD.92.063518}, and reveal the astrophysical and particle properties of the dark matter~\cite{PhysRevD.83.075002,Lee_2012,NAGAO2020100426}.

NEWAGE (NEw generation WIMP search with an Advanced Gaseous tracker Experiment) is a direction-sensitive WIMP dark matter search experiment using a low pressure gas micro time projection chamber ($\mu$TPC)~\cite{Miuchi_2003}. A direction-sensitive experiment needs to detect the direction of the recoil nuclei. Hence a TPC at low pressure gas and a readout device, $\mu$-PIC~\cite{Takada}, which is one variation of micro pattern gaseous detectors, are used. 
In 2015, an underground measurement was performed and the best directional constraint testing the forward-backward asymmetry in the nuclear recoil angular distribution was achieved~\cite{PTEPNEWAGE}. 
Then we increased the statistics by a factor of more than ten and the first 3d-vector directional dark matter search was performed~\cite{Yakabe2020}. 
However, a certain amount of radioactivity, which potentially contributed to the background, was later found inside the $\mu$-PIC. Therefore the surface material of the $\mu$-PIC, which was the dominant background source, was replaced with less radioactive material. This newly developed low-background $\mu$-PIC was called a low alpha-ray emission rate micro pixel chamber (LA$\mu$-PIC)~\cite{LAu-PIC}.
In this paper, the first results of the direction-sensitive dark matter search using the LA$\mu$-PIC are reported.

\section{NEWAGE-0.3b'' detector}
The NEWAGE-0.3b' detector was upgraded to NEWAGE-0.3b'' by replacing the readout device from the standard $\mu$-PIC to LA$\mu$-PIC.
Most part of the detector system is unchanged and 
we briefly summarize the structure of the NEWAGE-0.3b'' detector and its performance in this section.

\subsection{System}
\begin{figure}[h]
\centering
\includegraphics[width=5in]{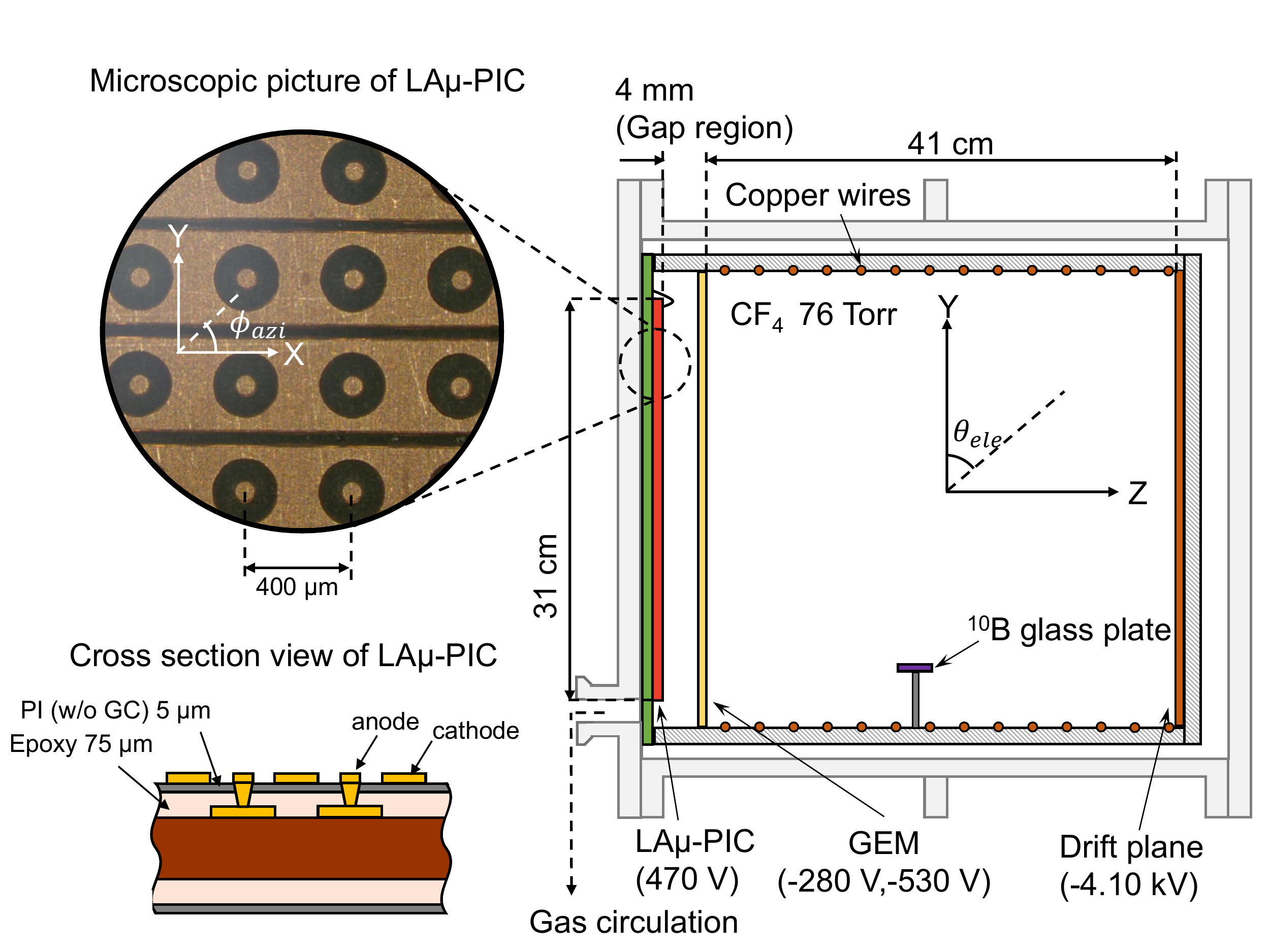}
\caption{Schematic drawings of the NEWAGE-0.3b'' detector. The left-top microscope picture shows the pixel electrode structure of the LA$\mu$-PIC. Left-bottom drawing is the cross section of the LA$\mu$-PIC. Surface material is combination of epoxy and polyimide without the glass cloth-sheet (GC). NEWAGE-0.3b'' chamber is shown in the right panel. LA$\mu$-PIC has a detection area of 31 $\times$ 31~cm$^{2}$ and the drift length is 41~cm. The gap region is defined as a volume between the LA$\mu$-PIC and the GEM. 
The $^{10}$B glass plate is set inside the TPC field cage for the energy calibration.}
\label{fig:detector}
\end{figure}

The NEWAGE-0.3b” detector is a low-pressure gaseous $\mu$TPC that is comprised of a LA$\mu$-PIC~\cite{LAu-PIC}, a gas electron multiplier (GEM~\cite{Sauli}) and a TPC cage. 
The schematic drawing is shown in Fig.~\ref{fig:detector}. 
The LA$\mu$-PIC was manufactured by Dai Nippon Printing Co. Lt.. 
It has 768 $\times$ 768 pixels with a pitch of 400~$\mu$m forming the detection area of 30.7~$\times$~30.7~cm$^{2}$. These electrodes are connected by 768 anode strips and 768 cathode strips. 
The anode and cathode strips of LA$\mu$-PIC are orthogonally formed and thus a two-dimensional position of a hit pixel can be known. 
The structure of the pixel electrode is same as that of a standard $\mu$-PIC~\cite{Takada}, while the material facing the detection volume was changed. 
Measurements of the U/Th contamination in the $\mu$-PIC components using a high purity germanium detector indicated that the glass cloth-sheet has a large amount of radioactive contamination. Therefore, a compound of epoxy and polyimide without the glass cloth-sheet, which is a factor of one hundred less contaminated by isotopes of $^{238}$U and $^{232}$Th, was chosen as a new surface material. Performance test showed that the newly developed $\mu$–PIC works as the standard $\mu$-PIC does. Details can be found in Ref.~\cite{HASHIMOTO2020}.
A GEM with an effective area of 32~$\times$~31~cm$^{2}$ is used as a first stage amplifier in order to obtain a sufficient gas gain while keeping a stable operation. 
The substrate of the GEM is a 100~$\mu$m thick liquid crystal polymer.
Cylindrical holes are formed using the laser etching technique. 
A hole size and pitch are 70~$\mu$m and 140~$\mu$m, respectively~\cite{TAMAGAWA}.
The TPC field cage, which is made of four plates of polyether ether ketone plastic plates, was installed in order to make a uniform electric field.
Copper wires with a spacing of 1~cm are placed on the side walls and chained by resistors. 
The TPC cage has a length of 41~cm. 
The vessel was filled with CF$_{4}$ at 76~Torr. CF$_{4}$ is chosen because of its small diffusion and a large cross section for the Spin-Dependent (SD) interaction of fluorine.  A gas circulation system with a cooled charcoal (TSURUMICOAL 2GS) of 100 g was installed in order to remove radons.

The charge signals are read out from the anode and cathode electrodes with  Amplifier-Shaper-Discriminator chips (SONY CXA3653Q~\cite{ASD_orito}).
These signals are divided into digitized and analog signals.
The digitized signals are sent to the FPGA-based encoding system and the hit-patterns are recorded with a clock of 100~MHz. In addition, the  time-over-thresholds (TOTs) of each strip are recorded. The track-length and direction are reconstructed using these information (see Ref.~\cite{Yakabe2020} for details).
Analog signals of the 768 cathode strips are grouped into four channels and each channel is then divided into two.
One of the divided signals is directly connected to a waveform digitizer (REPIC RPV160, 100~MHz), while
the other is attenuated by a factor of three and read by the waveform digitizer.
The waveforms are mainly used to determine the energy deposit of each event. 

The energy calibration is performed with alpha-rays generated in a $^{10}{\rm B}(n,\alpha)^{7}{\rm Li}$ reaction.
A glass plate with a thin $^{10}$B layer is installed inside the $\mu$TPC. The detector is irradiated with neutrons from a $^{252}$Cf fission source placed outside of the vessel and the neutrons are thermalized by polyethylene blocks. A continuous spectrum with a maximum edge at 1.5~MeV because of the thickness of the $^{10}$B layer is obtained.
The measured spectrum is fitted with simulated ones and the calibration factor converting the charge to energy is determined.

\subsection{Performance}
\label{sec:Performance}
\begin{figure}[!h]
\centering
\includegraphics[width=6in]{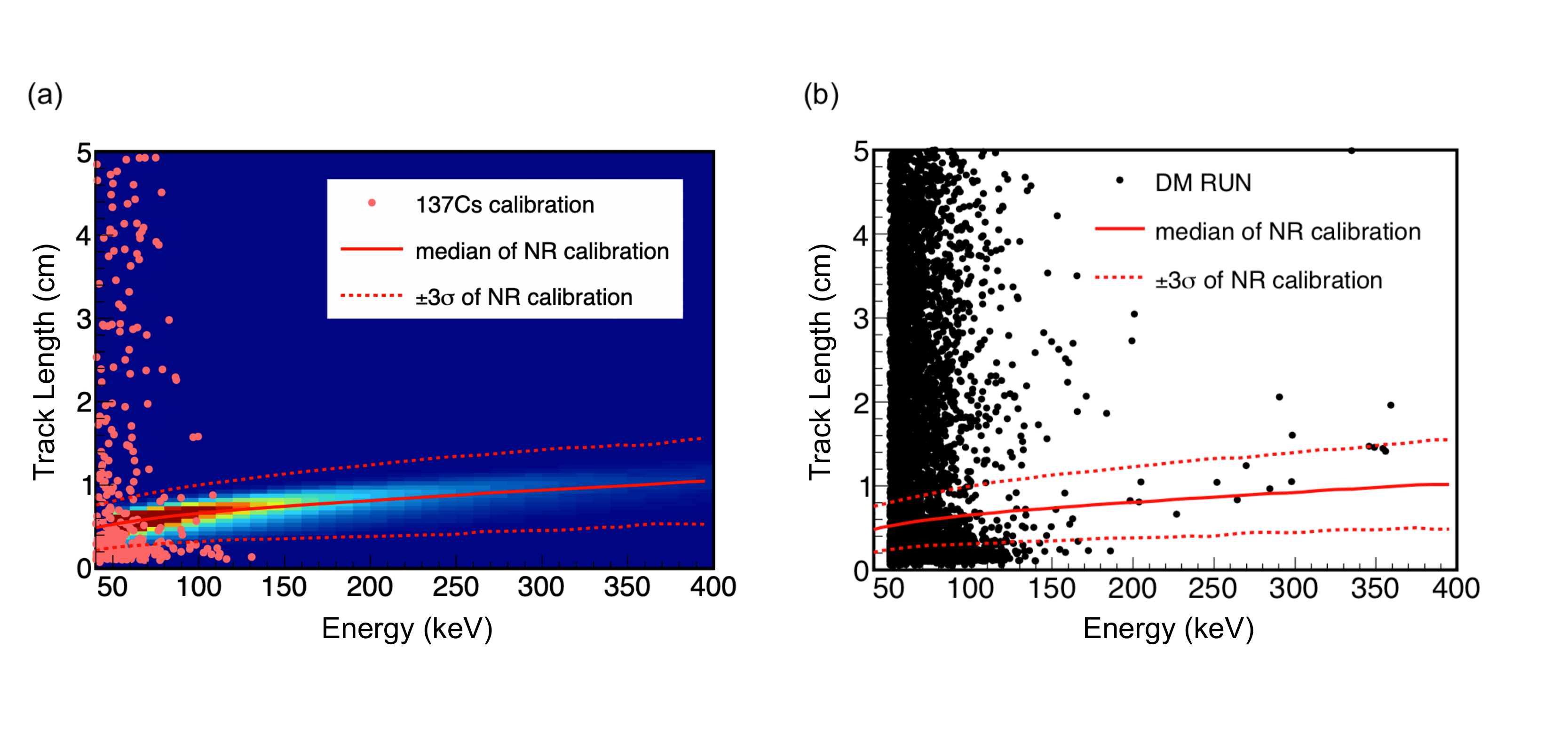}
\caption{Track-length vs. energy distributions. 
(a)~The gradation and red points represent the $^{252}$Cf neutron calibration data and the $^{137}$Cs electron calibration data, respectively.
The red solid and dotted lines indicate the median and $\pm$3$\sigma$ quantiles of the neutron calibration.
(b)~Scientific RUN data (DM RUN) after the fiducial cut.}
\label{fig:Track_Energy}
\end{figure}

\begin{figure}[!h]
\centering
\includegraphics[width=6in]{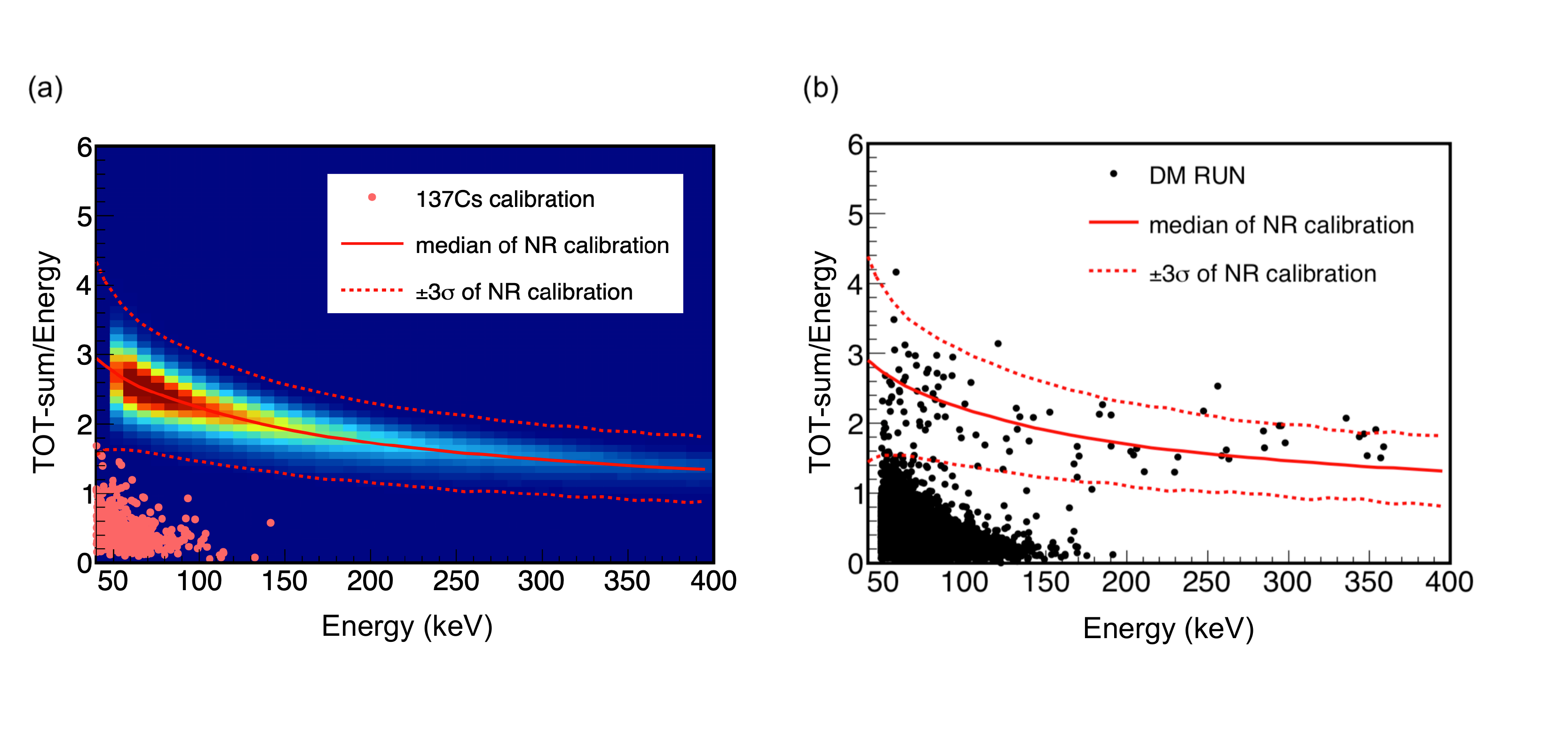}
\caption{TOT-sum/energy vs. energy distributions.
(a)~The gradation and red points represent the $^{252}$Cf neutron calibration data and the $^{137}$Cs electron calibration data, respectively.
The red solid and dotted lines indicate the median and $\pm$3$\sigma$ quantiles of the neutron calibration.
(b)~Scientific RUN data (DM RUN) after the fiducial cut.}
\label{fig:TOT}
\end{figure}

\begin{figure}[!h]
\centering
\includegraphics[width=6in]{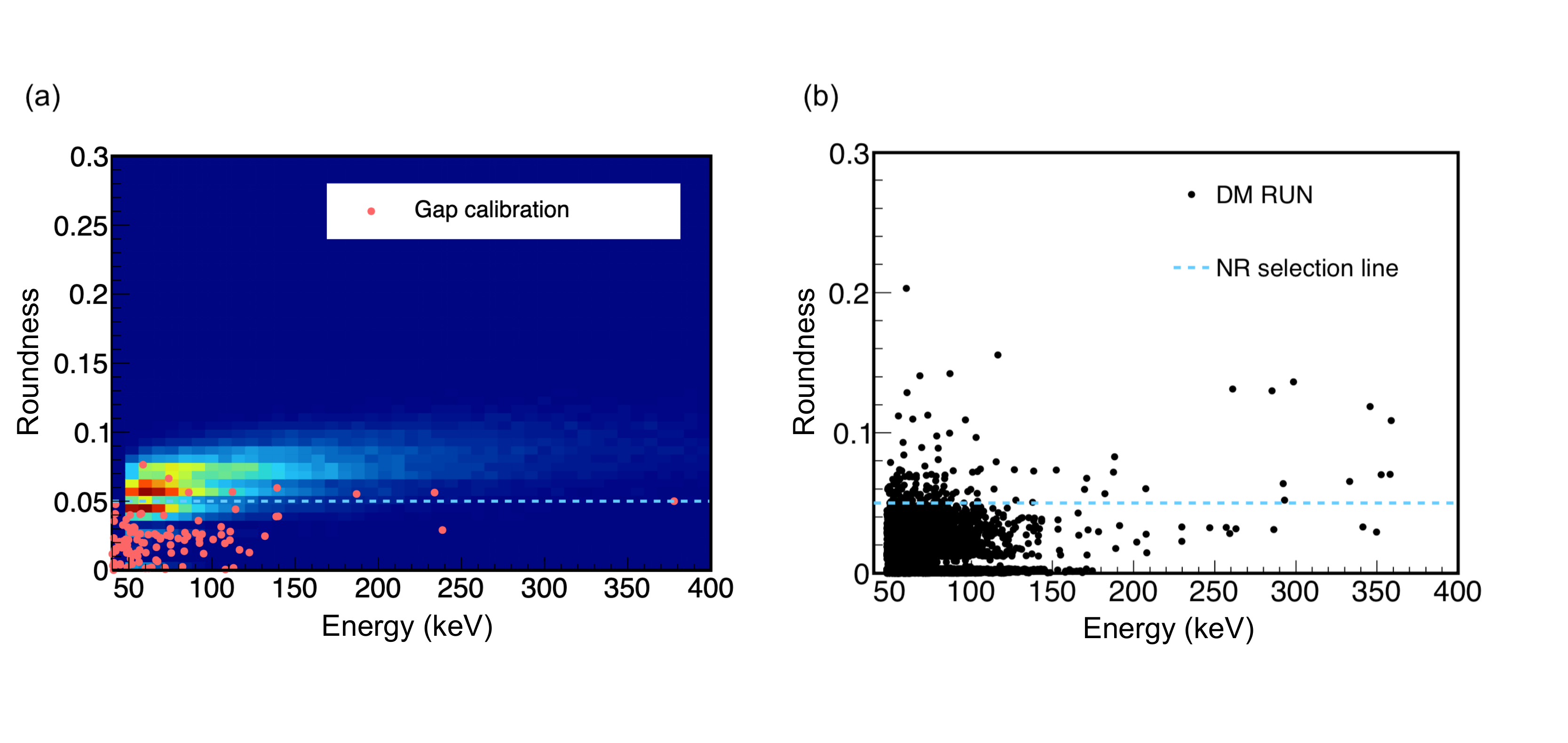}
\caption{Roundness vs. energy distribution.
(a)~The gradation and red points represent the $^{252}$Cf neutron calibration with a drift electric field and without a drift electric field, respectively.
The cyan dotted line (roundness$=0.05$) indicate the nuclear selection line.
(b)~Scientific RUN data (DM RUN) after the fiducial cut. }
\label{fig:Roundness}
\end{figure}

The following four analysis cuts were applied in order to select nuclear recoil events. Here the $^{252}$Cf neutron calibration data was taken without polyethylene blocks in order to induce the fast neutron-nuclear scatterings.
\begin{itemize}
\item Fiducial cut: The fiducial area of XY plane was defined as 28~$\times$~24~cm$^{2}$. The whole part of the track is required to be in the fiducial volume. This cut removes the charged particles from the wall and the $^{10}$B glass plate.
\item Length-Energy cut: The stopping powers of nuclei are larger than those of electrons. Hence the 
track-length vs. energy distribution can be used  to 
identify the nuclear recoil tracks.
Figure~\ref{fig:Track_Energy}~(a) shows the track-length vs. energy distributions for the $^{252}$Cf calibration (black points) and the $^{137}$Cs calibration (blue points). From the $^{252}$Cf calibration data, we determined the nuclear band by fitting with Gaussian function for every energy bin with a width of 10 keV. This cut removes electrons and alpha-rays.
\item TOT-Energy cut: Energy deposition on a single strip is stored as TOT. Since nuclear recoil events have larger energy losses than electron events, the sum of TOT (TOT-sum) for given energy tends to be large. In addition, since TOT-sum was expected to be linear with respect to the total energy deposit, a parameter defined by dividing the TOT-sum by energy was suitable for this purpose. Figure~\ref{fig:TOT}~(a) shows TOT-sum/energy vs. energy distributions for the $^{252}$Cf calibration (black points) and the $^{137}$Cs calibration (blue points). From the $^{252}$Cf calibration data, we determined the nuclear band by fitting with gaussian function every energy bin with a width of 10 keV. This cut rejects electron events.
\item Roundness cut: The roundness is defined as a reduced chi-square value of a linear fit to the track. While the self-triggering mode TPC is not able to measure the absolute z position, this parameter is known to show some correlation with the absolute z position. For instance, small z events have short drift lengths and thus, the effect of electron diffusion is negligible. The track information keeps its original shape and fitted with a straight line well. On the other hand, large z events are largely affected by electron diffusions, and the fitness to a straight line would be worse. In other words, these large z events have large roundness. Hence the roundness correlates with the absolute z position.
Especially, this parameter is useful to remove ``gap events''. The gap events are defined as events depositing their all energy in the gap region (Fig.~\ref{fig:detector}). These gap events are not amplified by the GEM and thus the measured charge is smaller by a factor of the GEM gain than those in the detection volume. In order to reproduce the gap events, we irradiated the detector with neutrons from a $^{252}$Cf fission source without a drift electric field. Figure~\ref{fig:Roundness} (a) shows the roundness vs. energy distribution for the gap events. The roundness of gap events (red points) was found to be small. The events with roundness $>$ 0.05 are selected in order to remove gap events.
\end{itemize}

\begin{figure}[!h]
\centering
\includegraphics[width=3in]{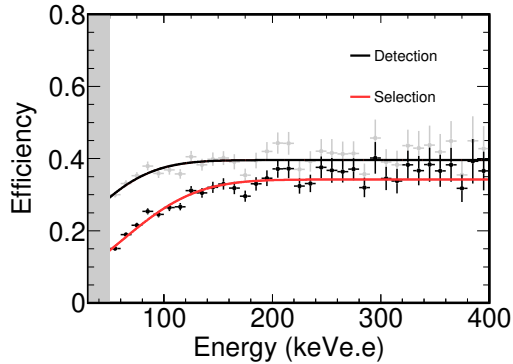}
\caption{Typical detection efficiency for nuclear recoil events.  
The gray and black points with errors are experimental data after the fiducial cut and the roundness cut, respectively.
The fitted black solid line is the detection efficiency in the fiducial volume.
The fitted red solid line is the detection efficiency after all event selections.}
\label{fig:Efficiency}
\end{figure}

The nuclear detection efficiency was evaluated by irradiating the $\mu$TPC with neutrons from a $^{252}$Cf fission source. In order to cancel the x, y and z position dependence of $\mu$TPC and evaluate overall efficiency, we needed homogeneous irradiation throughout the detection volume. We confirmed by Monte Carlo simulations that homogeneous irradiation in the TPC is made when $^{252}$Cf fission sources were set six positions. In the actual measurement, we set a $^{252}$Cf source at one of the six positions and took the data,  then move the source to another position. We performed the measurements six times in series and combined the data in the analysis to realize homogeneous irradiation.
The detection efficiency was calculated by comparing the measured energy and the simulated one.
The measured detection efficiencies of the nuclear recoil events are shown in Fig.~\ref{fig:Efficiency}.
The black and red lines are fitted lines of experimental data and indicate the detection efficiency when we applied only the fiducial cut and all event selections. The efficiency with all event selections was 14\% at 50 keV.

\begin{figure}[!h]
\centering
\includegraphics[width=6in]{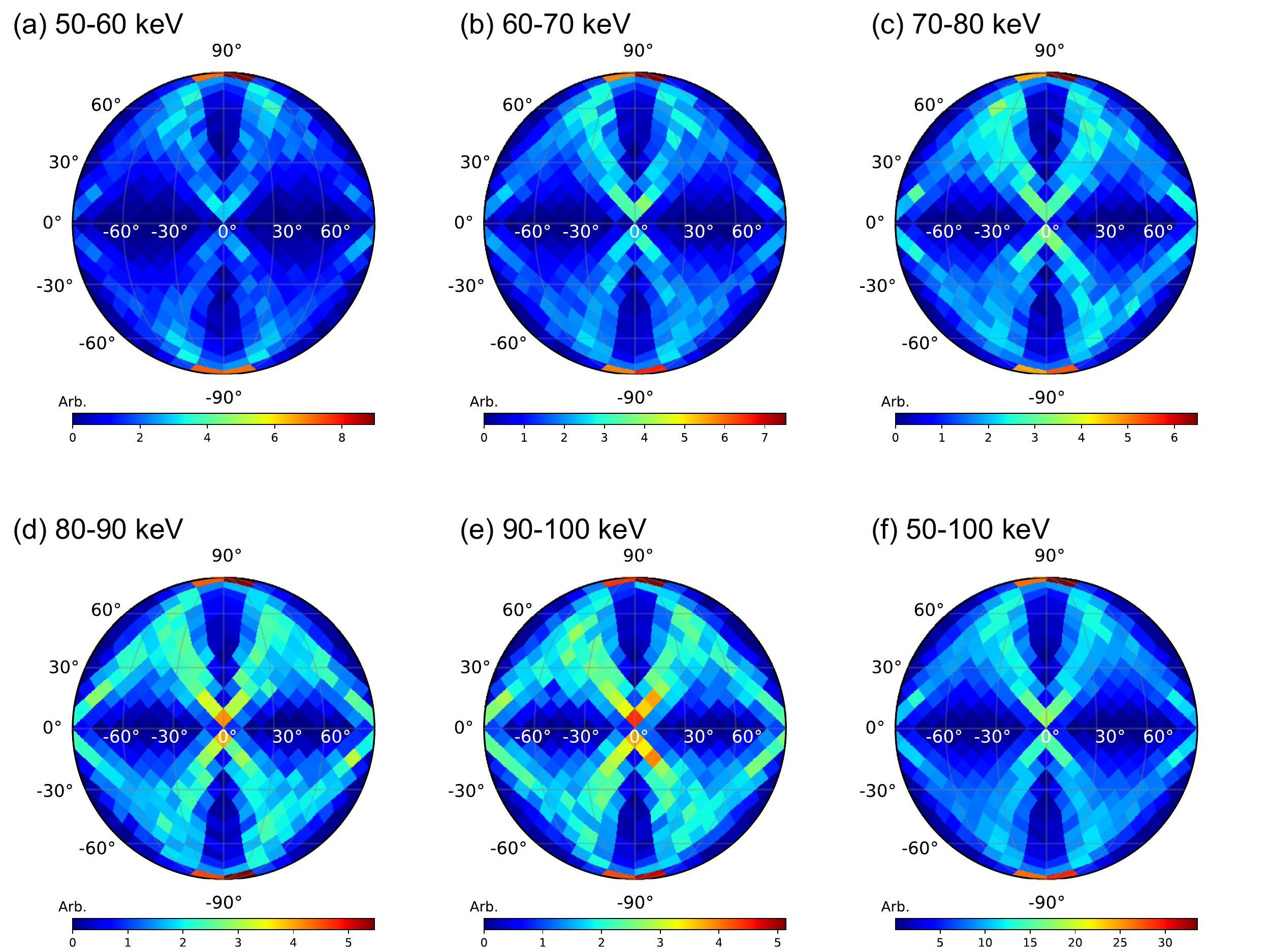}
\caption{Relative direction-dependent efficiencies in each energy region.
The axis with white and black label are the azimuth angle $\phi_{\rm{azi}}$ and the elevation angle $\theta_{\rm{ele}}$ in the detector coordinate, respectively. 
The x-axis and y-axis correspond to $(\phi_{\rm azi},\theta_{\rm ele})=(0,0)$ and $(\phi_{\rm azi},\theta_{\rm ele})=(90,0)$, respectively.
}
\label{fig:eff_direction_map_det}
\end{figure}

The $\mu$TPC has a non-isotropic response with regard to the nuclear recoil track direction because of the track-reconstruction algorithm.
Thus, we need to measure the relative direction-dependent efficiency of nuclear recoil in the energy range of 50-100 keV. 
Figure~\ref{fig:eff_direction_map_det} shows the measured distribution of the elevation angle $\theta_{\rm ele}$ and the azimuth angle $\phi_{\rm azi}$ in the detector coordinate. Although an ideal directional detector is expected to have a uniform distribution, our detector is not optimized yet in terms of the uniformity for the directional response.
The efficiency is low around XY plane, XZ plane and YZ plane in any energy region because of the poor track reconstruction along anode/cathode strips. 
In addition, there are higher efficiency areas along the diagonal lines from the direction of the anode and the cathode strips.
This is because the current tracking algorithm tends to recognize the diffused tracks to such directions. 
We evaluated the gamma rejection power, or the detection efficiency of electrons, by irradiating the detector with gamma-rays from a $^{137}$Cs source.
The gamma rejection power or the electron detection efficiency for the energy range of 50-60 keV  was $1.3^{+3.0}_{-1.1}\times10^{-6}$.

The angular resolution was evaluated using the neutron-nuclei elastic scatterings of same method in previous study~\cite{Nakamura_2012}. The angular $\theta$ is defined as the angle between the direction of the scattered nuclei and the neutron source.
We evaluated the angular resolution by the comparison of measured and simulated distributions of the recoil angle $\cos\theta$. The obtained angular resolution was 48.0$^{+6.8}_{-2.2}$ degree in the energy range of 50-100 keV. 
The expected forward-backward ratio of $|\cos\theta_{\rm{cygnus}}|$
for the 100~GeV/c$^{2}$ WIMP is 30\%. Here $\theta_{\rm{cygnus}}$ is defined as the angle between the WIMP-wind direction and the measured direction of the recoil nucleus.

\section{Experiment}
A directional dark matter search (RUN22) was carried in Laboratory B of the Kamioka Observatory (36.25’N, 137.18’E) located at 2700 m water equivalent underground. The LA$\mu$-PIC plane was placed vertically and the $z$-axis is aligned to the direction of S30$^{\circ}$E. 
The first sub RUN was performed from Jun. 6th 2018 to 24th Aug. 2018 and the second sub RUN was carried out from 20th Sep. 2018 to 14th Nov. 2018. 
The target gas is CF$_{4}$ at 76 Torr (0.1~bar) and the target mass in the fiducial volume of 28~$\times$~24~$\times$~41~cm$^{3}$ (28~L) is 10~g. 
The total live time is 107.6 days corresponding to an exposure of 1.1 kg$\cdot$days.

The energy calibration and the detection efficiency measurement were carried out every two weeks. The gas gain at the beginnings of the sub RUNs was about 1100 and time-dependent variation was observed due to the gas deterioration. The energy scale of the data was corrected considering the time-dependence of the gas gain. The energy resolution was $13.2 \pm 2.3\%$ above 50 keV.
The measured drift velocity at the beginnings of the sub RUNs was 9.6~cm/$\mu$s and the gas deterioration gave $+3.3\%$ and $-12\%$ of the maximum uncertainty. 

\begin{figure}[!h]
\centering
\includegraphics[width=6in]{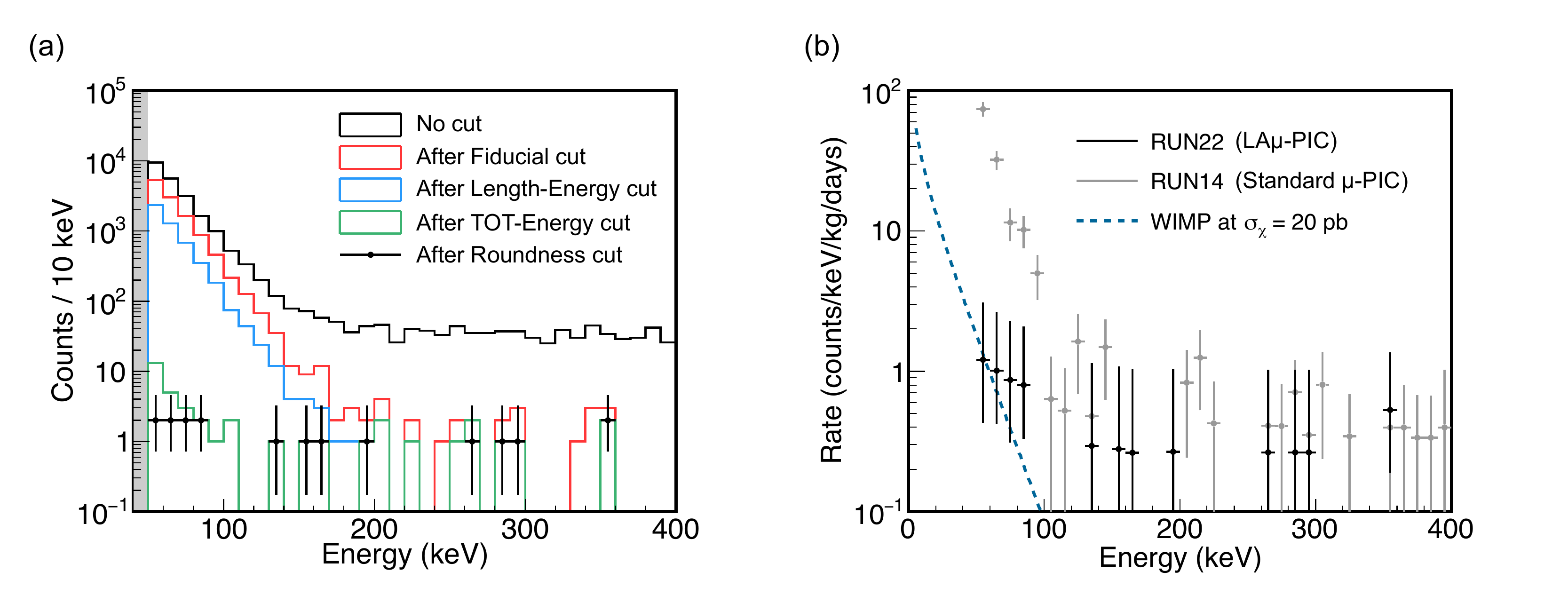}
\caption{(a) Energy spectra of the dark matter search at each selection step. The black, red, blue and green lines are no-cut energy spectrum, the one after the Fiducial cut, Length-Energy cut and TOT-Energy cut, respectively. The black points with error bars are the final event sample after the Roundness cut. 
(b) Final energy spectrum considering the detection efficiency. The black and gray points with error bars represent RUN22 (using the LA$\mu$-PIC) and RUN14 (using the standard $\mu$-PIC), respectively.
Error bars indicate statistic poisson errors. The blue dotted line shows the expected spectrum of the WIMP-nucleus scattering with the WIMP mass of 100~GeV/c$^{2}$, the WIMP-proton cross section of $\sigma_{\chi}=20~{\rm pb}$ and the energy resolution of 13.2\%.}
\label{fig:spectrum}
\end{figure}

Length-Energy, the TOT-Energy and the Roundness-Energy distributions 
after the fiducial cut for the dark matter search data 
are shown in Fig.~\ref{fig:Track_Energy} (b), \ref{fig:TOT} (b) and \ref{fig:Roundness} (b), respectively.
A large fraction of the events has long track lengths and small TOTs, which indicates that most of the measured events are electrons. These events are effectively reduced by analysis cuts introduced in Sec.~\ref{sec:Performance}.
Energy spectra at each cut stage are shown in Fig.~\ref{fig:spectrum} (a). The final event sample was reduced to 17 events in the energy region of 50-400~keV owing to analysis cuts using track information without any shields like Pb.
Since the directionality is lost in the energy below 50~keV because of the short tracks and the diffusion effect, the lower energy bound is set at 50~keV. 
The energy spectrum for the final sample unfolded by the nuclear detection efficiency 
is shown in Fig.~\ref{fig:spectrum}  (b) together with one of the previous results RUN14~\cite{PTEPNEWAGE} using the standard $\mu$-PIC.
The main background of RUN14 in 50-100~keV are alpha-rays radiated from the surface material of the standard $\mu$-PIC.
These backgrounds were reduced in RUN22 by about factor 10 thanks to the LA$\mu$-PIC whose surface material is less contaminated with $^{238}$U and $^{232}$Th.
It is demonstrated that the LA$\mu$-PIC works as expected to reduce the alpha-ray backgrounds.

\begin{figure}[!h]
\centering
\includegraphics[width=3.5in]{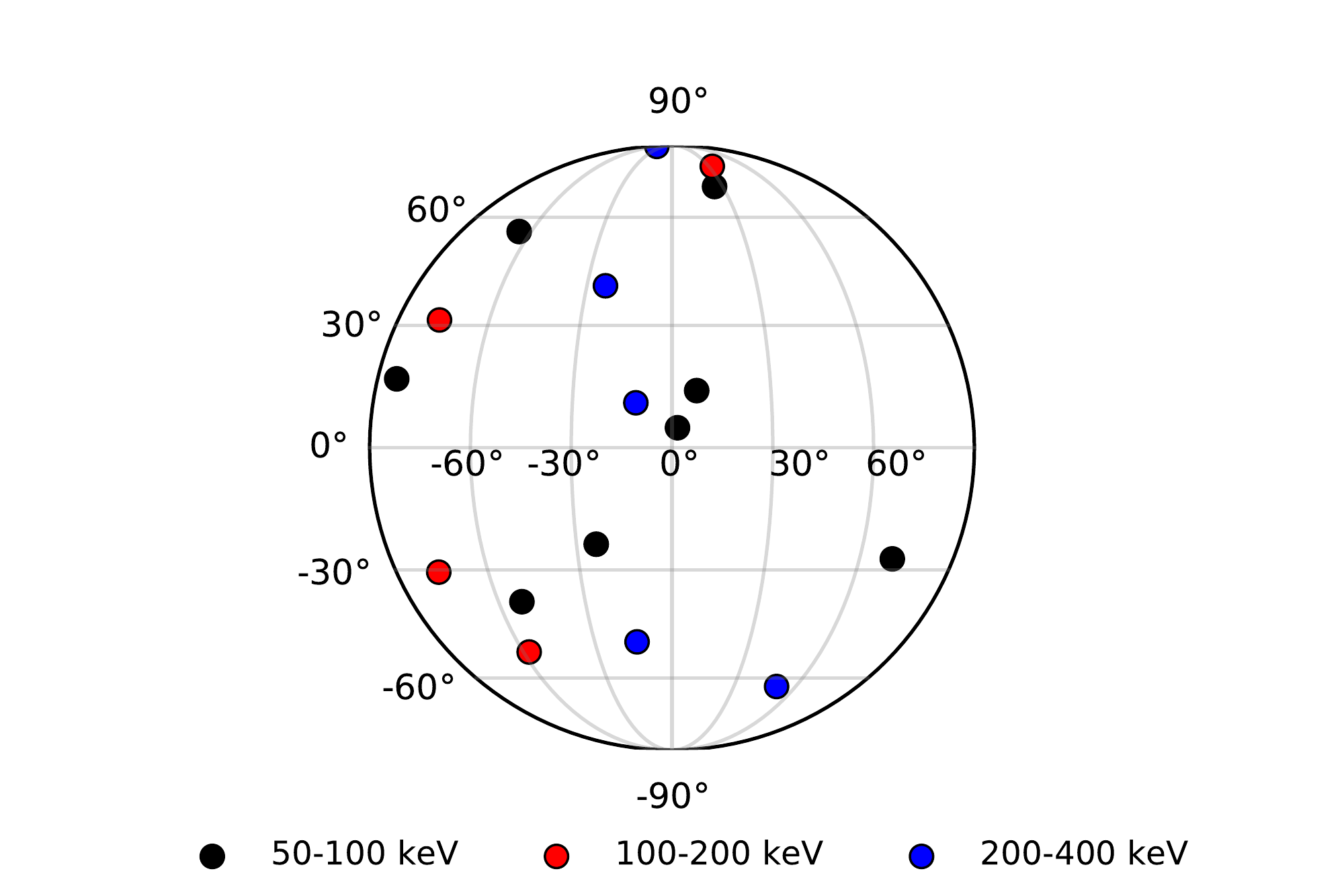}
\caption{Skymap of the final sample on the detector coordinate.
The black, red and blue points show the energy region of 50-100 keV, 100-200 keV and 200-400 keV, respectively.}
\label{fig:skymap}
\end{figure}

\begin{figure}[!h]
\centering
\includegraphics[width=5.5in]{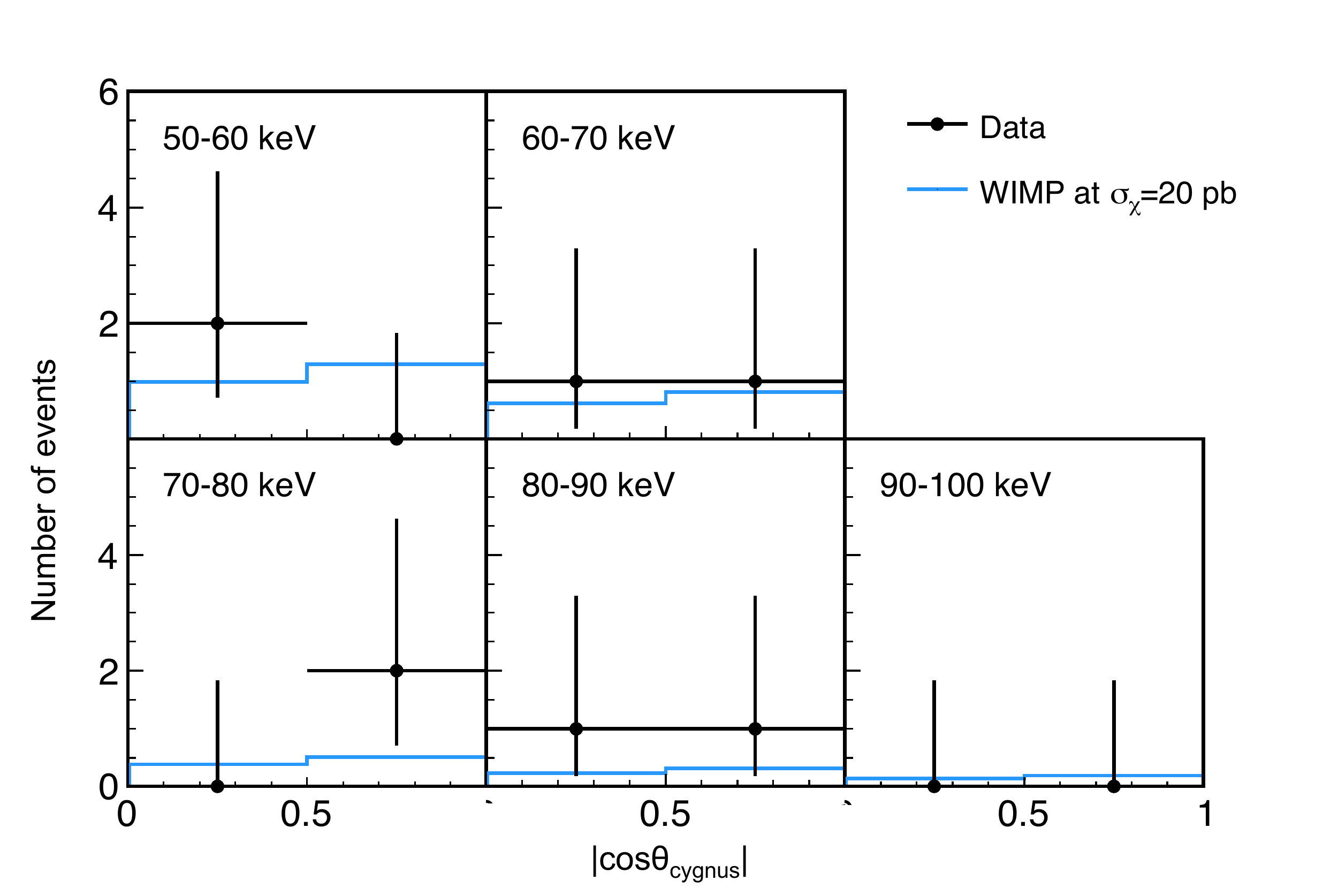}
\caption{Measured and expected $|\rm{cos}\theta_{\rm{cygnus}}|$ distribution for each energy range.
The black points with errors are measured data. The cyan lines show the WIMP expected signal of the WIMP-nucleus scattering with the WIMP mass of 100~GeV/c$^{2}$ and the WIMP-proton cross section of $\sigma_{\chi}=20~{\rm pb}$.}
\label{fig:cos_data}
\end{figure}

Figure~\ref{fig:skymap} shows the skymap of the final sample on the detector coordinate.
We calculated the nuclear recoil distribution $|\cos\theta_{\rm{cygnus}}|$ for the energy region of 50-100 keV in order to evaluate the forward-backward asymmetry. 
Figure.~\ref{fig:cos_data} shows the measured and expected $|\cos\theta_{\rm{cygnus}}|$ distribution binned into two for each energy range.
As a model-independent interpretation of these results, we first set upper limits on the forward-backward asymmetry parameters based on the raw counting numbers in each energy bin. The forward-backward asymmetry parameter is defined as the ratio of the first and second bins. The 90\% confidence level upper limits of these values in 50-60~keV, 60-70~keV, 70-80~keV and 80-90~keV are 2.3, 3.9, 5.3 and 3.9, respectively.
The systematic uncertainties of the expected rate for the WIMP were summarized in Table~\ref{tab:systematic_uncertainties}.
The angular resolution gives the dominant systematic uncertainty, which impacts the shape of the nuclear recoil distribution, and is considered in the following statistic test.


\begin{table}[!h]
\centering
\caption{Systematic uncertainties of the expected rate for the WIMP mass $m_{\chi}=100$ GeV/c$^{2}$.}
\begin{tabular}{lcc} \hline
   Source                              & $\cos\theta$ range & Relative uncertainty (\%) \\ \hline
   Energy resolution                   & [ 0, 1 ]                    & $<0.1$ \\
   \multirow{2}{*}{Drift velocity}     & [ 0, 0.5 ] & $<0.2$ \\
                                       & [ 0.5, 1 ] & $<0.2$ \\
   \multirow{2}{*}{Angular resolution} & [ 0, 0.5 ] & +5.4 $-2.2$ \\
                                       & [ 0.5, 1 ] & +1.7 $-4.2$ \\
  \hline
\end{tabular}
\label{tab:systematic_uncertainties}
\end{table}

\section{Results}
In order to obtain a possible anisotropic $|\rm{cos}\theta_{\rm{cygnus}}|$ distribution, a binned likelihood-ratio method was used \cite{PhysRevD.67.012002}.
The minimized statistic value $\chi^{2}$ was defined as,
\begin{equation}
\label{eq:chi_square}
\chi^{2} = 2 \sum_{i=0}^{n} \biggl[ ( N_{i}^{\rm{exp}} - N_{i}^{\rm{data}} ) + N_{i}^{\rm{data}} {\rm{ln}}\biggl(\frac{ N_{i}^{\rm{data}}}{N_{i}^{\rm{exp}}}\biggr) \biggr] + \alpha^{2},
\end{equation}
where the subscript $i$ is the bin number of $|\cos\theta_{\rm{cygnus}}|$ distribution, $N_{i}^{\rm{data}}$ is
the measured number of events and $N_{i}^{\rm{exp}}$ is the expected number of events.
A nuisance parameter $\alpha$ $(=\xi/\sigma_{\kappa})$ was introduced to consider the systematic uncertainty of the angular resolution $\sigma_{\kappa}$.
A possible angular-resolution shift is $\xi$. 
The expected event number $N_{i}^{\rm{exp}}$ was obtained by a signal Monte Carlo simulation.
It depends on the WIMP mass $m_{\chi}$, the WIMP-proton cross section $\sigma_{\chi-p}$ and the astrophysical parameters.
In addition, the nuclear quenching factor and the detector responses were considered.
Here the nuclear quenching factor was simulated by SRIM~\cite{SRIM}, which represented the experimental alpha-ray data of the previous experiment~\cite{NISHIMURA2009185}.
The energy bin was divided into 10~keV/bin considering the energy resolution.
Since each energy bin had low statistics, the measured $|\cos\theta_{\rm{cygnus}}|$ were binned into 2-bin.

\begin{figure}[!h]
\centering
\includegraphics[width=3in]{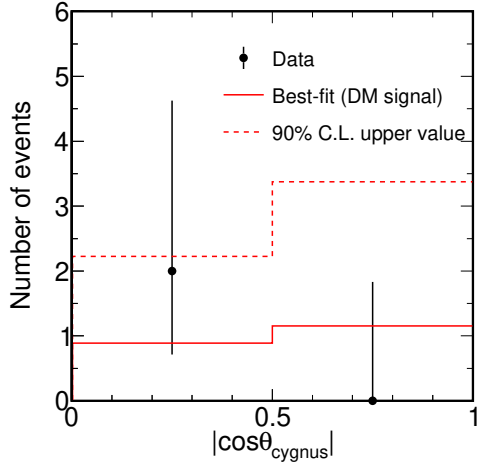}
\caption{The $|\cos\theta_{\rm{cygnus}}|$ distribution in the energy range of 50-60~keV.
The black points show observed data.
The solid and dotted red lines are the simulated distribution using best-fit values and excluded values of 90\% C.L., respectively.}
\label{fig:cos_fit_result}
\end{figure}

\begin{figure}[!h]
\centering
\includegraphics[width=4.0in]{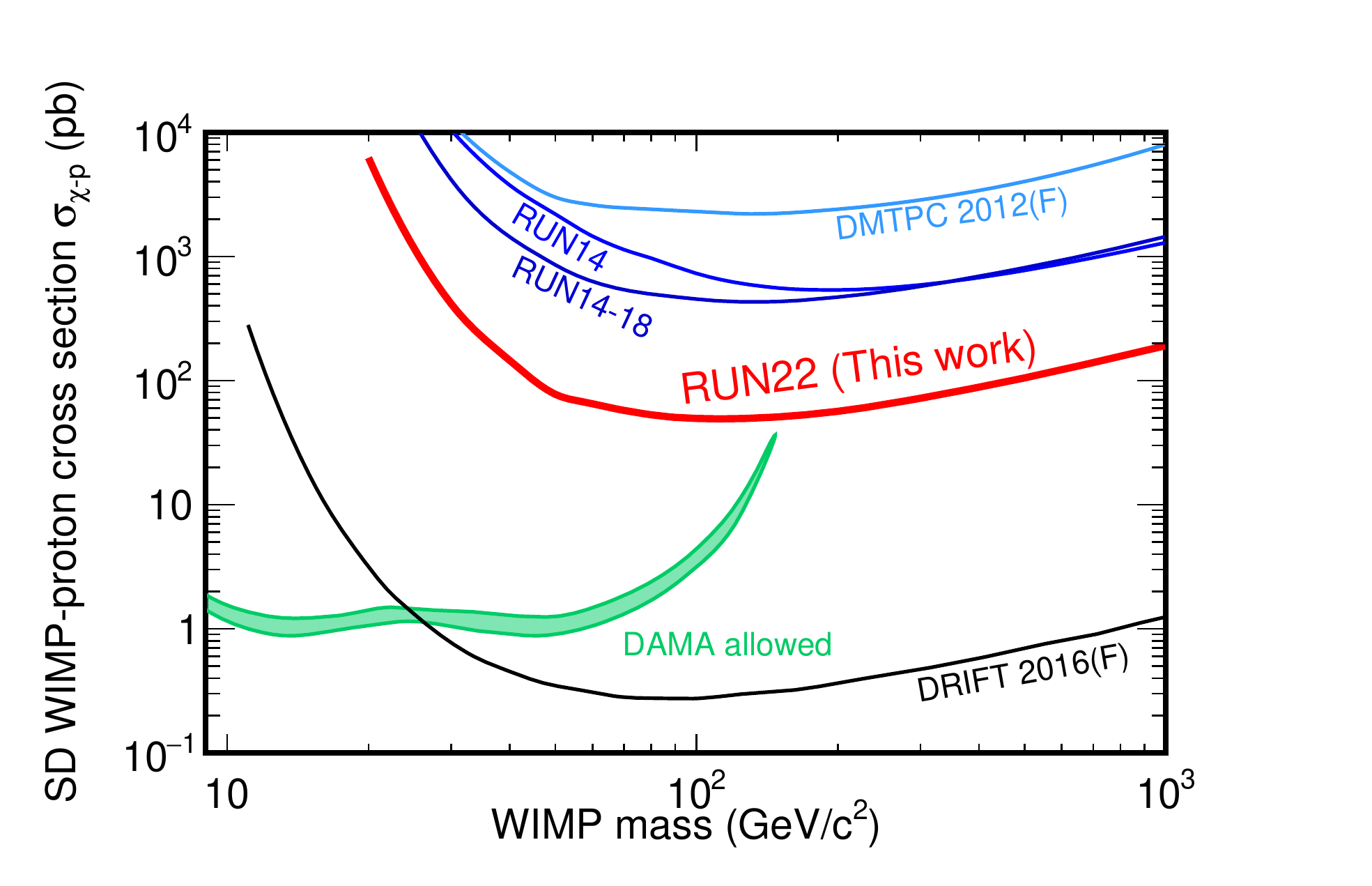}
\caption{90\% C.L. upper limits on the SD WIMP-proton cross section as a function of the WIMP mass.
The red thick solid line is the result of the directional method in this work.
The blue solid lines of RUN14 and RUN14-18 are our previous results~\cite{PTEPNEWAGE,Yakabe2020}.
The light blue and black solid lines show the results from the directional analysis of DMTPC~\cite{DMTPC} and the conventional analysis without the directional sensitivity of DRIFT~\cite{BATTAT20151}, respectively.
Allowed region from DAMA/LIBRA experiment~\cite{PhysRevD.70.123513} is shown by the green area.}
\label{fig:limit}
\end{figure}

The measurement data was fitted by minimizing $\chi^{2}$ using an anisotropic WIMP model. Here, no background (BG) was added  in order to prevent any uncertainty of the BG distribution.
The WIMP-proton cross section $\sigma_{\chi-p}$ and the nuisance parameter $\alpha$ were treated as fitting parameters.
The minimum $\chi^{2}$ value for the 50-60~keV bin was 3.3 where $\sigma_{\chi-p}$ and $\alpha$ were 18.5~pb and 0.12, respectively.
Figure~\ref{fig:cos_fit_result} shows the measured $|\cos\theta_{\rm{cygnus}}|$ distribution in the energy region of 50-60~keV along with the expected one using best-fit values.
In order to calculate the p-value, we made a $\chi^{2}$ distribution of an isotropic BG model and an anisotropic WIMP model from dummy samples.
One thousand dummy samples were produced by Monte Carlo simulations and $\chi^{2}$ value of each dummy sample was calculated.
P-values for both of the WIMP and BG model were 3.3\%.
Hence we cannot claim the detection of WIMP dark matter with sufficient significance from the observed data.
This is a natural result because of the large statistic error and the small expected anisotropic ratio.
Since no significant amplitude was found, a 90\% confidence level (C.L.) upper limit was set on the SD cross section.
The 90\% C.L. upper limit on the SD cross section was obtained as 50~pb for 100~GeV WIMPs.
Figure~\ref{fig:limit} shows the 90\% C.L. upper limit on the SD WIMP-proton cross section as a function of the WIMP mass.
This result marked a new best sensitivity record of the SD WIMP search with the direction-sensitive method.
This result improved the constraint by about 15 times compared to the previous result of RUN14-18.
This improvement owes to the surface background reduction of the $\mu$-PIC detector.

\section{Discussion}
One of the milestones of the NEWAGE is to search an allowed region from DAMA experiment~\cite{PhysRevD.70.123513} with the directional method.
More than two orders magnitude of improvements is required in order to cover the entire DAMA region. Since the sensitivity is limited by the remaining backgrounds, we investigated the origin of these backgrounds in RUN22 using Geant4~\cite{Geant4} simulation.

The main internal background candidates are alpha-rays coming from the decay of radons and the LA$\mu$-PIC surface.
The alpha-rays cannot be discriminated from nuclear recoil events around 50~keV in NEWAGE-0.3b'' detector by analysis cuts. 
Hence screening the material of the TPC and the LA$\mu$-PIC is needed.
A broad peak is observed in the energy spectrum around 6~MeV corresponding to the $^{220}$Rn and $^{222}$Rn decay during the dark matter search~\cite{IkedaDron}. Estimated contamination of $^{220}$Rn and $^{222}$Rn are 5.7~$\pm$~0.3~mBq/m$^{3}$ and (5.3~$\pm$~2.1)~$\times$~10$^{-1}$~mBq/m$^{3}$, respectively. 
In Geant4 simulation, we generated alpha-rays from decay chains of $^{220}$Rn and $^{222}$Rn inside the TPC according to the branching ratio and estimated the number of events or rate with the analysis cut same to RUN22.
The expected number of events (rate) due to $^{220}$Rn and $^{222}$Rn contamination in the low energy region of 50-60~keV were $(6.1\pm0.7)\times10^{-1}$~(5.5~$\times$~10$^{-2}$~dru\footnote{dru$=$counts/sec/keV/kg}) and $(5.3\pm2.2)\times10^{-2}$ (4.7~$\times$~10$^{-3}$~dru), respectively.
The remaining alpha-ray emission rate of the LA$\mu$-PIC was (2.1~$\pm$~0.5)~$\times$~10$^{-4}$~alpha/cm$^{2}$/hr~\cite{HASHIMOTO2020} and the dominant component was 5.3 MeV alpha-rays from $^{210}$Po decay. The expected number of events (rate) in 50-60~keV region from this background was less than $1.2\times10^{-1}$ (1.1~$\times$~10$^{-2}$~dru).

Ambient gamma-rays and neutrons from rocks in the mine are two of the main components of the external background.
Contributions of the cosmic-ray muons are negligible compared with ambient gamma-rays and neutrons.
Measured ambient gamma-rays flux 
in the Laboratory B  ~\cite{Nishimura} was used for the estimation.
The expected counts (rate) in the energy range of 50-60~keV was $1.5\pm1.5$ (1.4~$\times$~10$^{-1}$~dru).
The ambient neutrons flux was measured using a $^{3}$He proportional counter and an energy spectrum of ambient neutrons produced by $(\alpha,n)$ reactions and spontaneous fission was predicted~\cite{Mizukoshi}. The expected counts (rate) in the energy range of 50-60~keV is $(3.5\pm0.9)\times10^{-1}$ (3.1~$\times$~10$^{-2}$~dru).
The expected number of background events in the energy region of 50-60~keV is summarized in Table~\ref{tab:dru_bgs}. The result without roundness-cut is also shown in order to confirm the alpha-ray backgrounds from the LA$\mu$-PIC surface, directly. The measured number of events are in good agreement with the predicted one within errors in both with and without roundness-cut.

The $^{222}$Rn backgrounds were reduced by the gas circulation system with cooled charcoal and their contribution was found to be negligible.
Ambient gamma-rays and ambient neutrons contribute some part of the backgrounds. These backgrounds can be reduced by the external shields comprising of materials like lead (Pb) and water (H$_{2}$O). 
On the other hand, internal backgrounds of $^{220}$Rn and the LA$\mu$-PIC surface would remain with the external shields and they would be dominant backgrounds.
A straightforward way of reducing these backgrounds is to replace the detector components with radiopure materials. A $\mu$-PIC with a further background reduction is being developed. Another approach to reduce the background is to detect the absolute $z$ position of the events.
The majority of the remaining backgrounds are known to locate at low $z$ (around the LA$\mu$-PIC and the GEM) and high $z$ position (around the drift plate). A discovery of minority carriers in CS$_{2}$ + O$_{2}$ gas mixtures by the DRIFT group opened the potential of an absolute $z$ measurement in self-triggering TPCs~\cite{BATTAT20151}. We have recently demonstrated a three-dimensional tracking with a spatial resolution of 130~$\mu$m using the $\mu$-PIC in a negative ion gas SF$_{6}$. Simultaneously, the absolute $z$ coordinate was determined with a location accuracy of 16~mm~\cite{Ikeda2020}. The negative ion gas TPC enables us to reduce these backgrounds using the $z$ fiducialization, effectively. Thus the use of the negative ion gas TPC is another promising approach to reduce the surface alpha-ray backgrounds. It should be noted that the surface background will be one of the ultimate background sources even after significant efforts of material selection since radioactive isotopes can be embedded by the decays of radons in a normal atmosphere even after the production.
Thus it is important to take both possible ways to reduce the background  
so as to start investigating the 
DAMA region and further searches.

\begin{table}[!h]
\centering
\caption{Summary of the expected numbers of background events and measured numbers in the energy region of 50-60~keV.}
\begin{tabular}{lcc} \hline 
  Source                         & w/ roundness                 & w/o roundness  \\ \hline
  Ambient gamma-rays             & $1.5\pm1.5$                  & $4.6\pm2.7$                   \\
  Ambient neutrons               &$(3.5\pm0.9)\times10^{-1}$    & $(4.8\pm1.2)\times10^{-1}$    \\
  $^{222}$Rn                     & $(5.3\pm2.2)\times10^{-2}$   & $(8.6\pm3.5)\times10^{-2}$    \\
  $^{220}$Rn                     & $(6.1\pm0.7)\times10^{-1}$   & $1.1\pm0.1$                   \\
  LA$\mu$-PIC surface            & $<1.2\times10^{-1}$          & $9.1\pm2.3$                   \\ \hline
  Total background               & $2.5\pm1.5$                  & $15\pm3.5$                    \\ \hline
  Measurement                    & $2.0\pm1.4$                  & $12\pm3.5$                    \\
  \hline
\end{tabular}
\label{tab:dru_bgs}
\end{table}

\section{Conclusion}
We developed a low background $\mu$TPC detector, NEWAGE-0.3b'', for the directional dark matter search with an LA$\mu$-PIC, which is a two-dimensional tracking gaseous detector made of low radioactive materials.
A directional dark matter search in the Kamioka Observatory was carried out in 2018.
Total exposure was 1.1~kg$\cdot$days and the number of the observed events in the energy region of 50-60~keV was two which was consistent with 2.5 events of the expected background.
No significant forward-backward asymmetry of a WIMP signal was found, 
therefore we derived a 90\% confidence level upper limits on the SD WIMP-proton cross section of 50~pb for 100~GeV/c$^{2}$ WIMPs.
We improved the constraint of the previous result~\cite{PTEPNEWAGE} by a factor of 15 and marked the best direction-sensitive limit.

\section*{Acknowledgment}
This work was partly supported by JSPS (Japan Society for the promotion of Science) KAKENHI (Grant-in-Aids for Scientific Research) (grant nos. 16H02189, 19684005, 23684014 and 26104005, 19H05806), JSPS Bilateral Collaborations (Joint Research Projects and Seminars) program,
ICRR Joint-Usage, Program for Advancing Strategic International Networks to Accelerate the Circulation of Talented Researches (R2607), and JSPS Research Fellow ( grant nos. 17J03537).


\bibliographystyle{ptephy}
\bibliography{sample}
%

\end{document}